\documentclass[twocolumn, pra, aps, superscriptaddress,longbibliography,showpacs,amsmath,amssymb,floatfix]{revtex4-1}                                                                                       \usepackage{graphicx}
\usepackage{amssymb}
\usepackage{amsmath}
\usepackage{epsfig}
\usepackage{color}
\usepackage{tabu}
\usepackage{mathtools}
\usepackage[colorlinks,linkcolor=blue,anchorcolor=blue,citecolor=blue,urlcolor=blue]{hyperref}
\usepackage{physics}
\usepackage{float}
\usepackage{diagbox}

\begin{document}

\title{Scaling of finite size effect of $\alpha$-R\'enyi entropy in disjointed intervals under dilation}
\author{Long Xiong}
\affiliation{CAS Key Laboratory of Quantum Information, University of Science and Technology of China, Hefei, 230026, China}
\author{Shunyao Zhang}
\affiliation{CAS Key Laboratory of Quantum Information, University of Science and Technology of China, Hefei, 230026, China}
\author{Guang-Can Guo}
\affiliation{CAS Key Laboratory of Quantum Information, University of Science and Technology of China, Hefei, 230026, China}
\affiliation{Synergetic Innovation Center of Quantum Information and Quantum Physics, University of Science and Technology of China, Hefei, Anhui 230026, China}
\affiliation{CAS Center For Excellence in Quantum Information and Quantum Physics,  University of Science and Technology of China, Hefei, Anhui 230026, China}
\author{Ming Gong}
\email{gongm@ustc.edu.cn}
\affiliation{CAS Key Laboratory of Quantum Information, University of Science and Technology of China, Hefei, 230026, China}
\affiliation{Synergetic Innovation Center of Quantum Information and Quantum Physics, University of Science and Technology of China, Hefei, Anhui 230026, China}
\affiliation{CAS Center For Excellence in Quantum Information and Quantum Physics,  University of Science and Technology of China, Hefei, Anhui 230026, China}
\date{\today }

\begin{abstract}
The $\alpha$-R\'enyi entropy in the gapless models have been obtained by the conformal field theory, 
which is exact in the thermodynamic limit. However, the calculation of its 
finite size effect (FSE) is challenging. So far only the FSE in a single interval in the XX model has been 
understood and the FSE in the other models and in the other conditions are totally unknown. 
Here we report the FSE of this entropy in disjointed intervals $A = \cup_i A_i$ under a uniform dilation 
$\lambda A$ in the XY model, showing of a universal scaling law as 
	\begin{equation*}
	\Delta_{\lambda A}^\alpha = 
\Delta_A^\alpha \lambda^{-\eta} \mathcal{B}(A, \lambda),
	\end{equation*}
		where $|\mathcal{B}(A, \lambda)| \le 1$ is a bounded function 
and $\eta = \text{min}(2, 2/\alpha)$ when $\alpha < 10$. We verify this relation in the phase boundaries of the XY model, 
in which the different central charges correspond to the physics of free Fermion and free Boson models. We find
that in the disjointed intervals, two FSEs, termed as extrinsic FSE and intrinsic FSE, are required to fully account
for the FSE of the entropy. Physically, we find that  only the edge modes of the correlation matrix localized at the open 
ends $\partial A$ have contribution to the total entropy and its FSE. Our results provide some incisive insight into the 
entanglement entropy in the many-body systems. 
\end{abstract}
	
\maketitle

\section{Introduction}

\begin{figure}[t]
    \centering
    \includegraphics[width=0.48\textwidth]{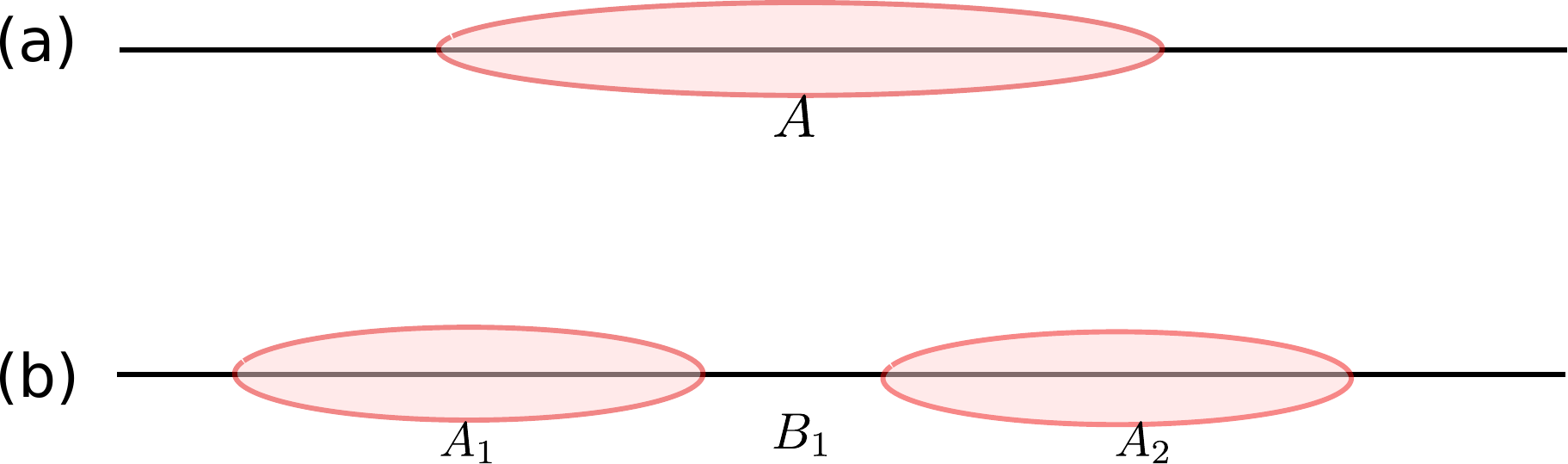}
    \caption{(a) Entanglement entropy in a single interval $A$; and (b) Entanglement entropy in two disjointed intervals 
	$A_1$ and $A_2$ separated by an interval $B_1$. The configuration in (b) can be generalized to disjointed 
	intervals; see results in Fig. \ref{fig-fig5} and Fig. \ref{fig-fig6}. The correlation between different intervals 
	will be calculated by the correlation functions $\mathcal{G}(x)$. }
	\label{fig-fig1}
\end{figure}

Entanglement has played a more and more important role in quantum information and many-body physics. A large
number of investigations have shown that 
the ground state of the gapped and gapless phases will have totally different entanglement entropies. For a regime $A$ 
(see Fig. \ref{fig-fig1} (a)), we can denote the reduced density matrix as $\rho_A$, then the Shannon entropy can be 
calculated using $S_A = -\text{Tr}(\rho_A \ln \rho_A)$. In the gapped phase, its entropy satisfies the area law 
\cite{eisert2010colloquium, verstraete2006criticality, srednicki1993entropy, zeng2019quantum, Swingle2016}. 
\begin{equation}
	S_A = \tilde{\alpha} \partial A -\tilde{\gamma} \sim L^{d-1}. 
	\label{eq-Area}
\end{equation}
However, in the gapless phase, it satisfies a different area law with logarithmic correlation as 
\cite{gioev2006entanglement, herdman2017entanglement}
\begin{equation}
S_A \sim L^{d-1} \ln L.
	\label{eq-SALog}
\end{equation}
In the above two equations, $L$ is the system size and $d$ is its system dimension.
When $d = 1$, it yields the Logarithm divergence of the entropy with
the increasing of system size (see below). The similar features may also be found for their low lying excited states
\cite{masanes2009area,alba2009entanglement,alcaraz2011entanglement,berganza2012entanglement}. By generalizing this concept 
in terms of $\alpha$-R\'enyi entropy, one find that in the one dimensional gapless phase \cite{renyi1961measures,jin2004quantum}
\begin{equation}
	S_A^\alpha = {1\over 1-\alpha} \log_2 \text{Tr} \rho_A^\alpha = 
	{c + \bar{c} \over 12(1+\alpha)} \log_2 L + s_0^\alpha + \Delta_A^\alpha,
	\label{eq-SA}
\end{equation}
where $c$ and $\bar{c}$ are the holomorphic and antiholomorphic central charges respectively \cite{calabrese2009entanglement,Hozhey1994}, $s_0$ is a non-universal constant 
and $\Delta_A^{\alpha}$ is its finite size effect (FSE), satisfying 
\begin{equation}
\lim_{L \rightarrow \infty} \Delta_A^{\alpha} = 0,
\label{eq-DeltaL}
\end{equation}
by its definition. The expression of $\alpha$-R\'enyi entropy has been examined numerically in some of the solvable 
models \cite{refael2004entanglement, amico2008entanglement,horodecki2009quantum, vidal2003entanglement,latorre2003ground,
jin2004quantum,levine2004entanglement,franchini2007renyi}, which can be more rigorously obtained by the conformal field theory 
(CFT) \cite{Calabrese_2004,peschel2009reduced,calabrese2009entanglement,latorre2009short}. Since the gapped phase and gapless 
phases have totally different entanglement properties, these features are used to diagnose the phase transitions in some of
the many-body models \cite{chhajlany2007entanglement,apollaro2016entanglement,orus2008equivalence,jia2008entanglement,
zozulya2009entanglement,lundgren2014momentum,dumitrescu2017scaling}. 

The scaling laws of the FSE in Eq. \ref{eq-DeltaL}, which in the gapped and gapless phases should exhibit totally different 
behaviors, are the major concern of this manuscript. To date, it has been rarely investigated. In the XX model with free
fermions \cite{vidal2003entanglement,latorre2003ground,casini2007c,jin2004quantum}, it 
has been calculated using the Jin-Korepin (JK) approach \cite{jin2004quantum,calabrese2010universal}, yielding a extremely 
complicated polynomial of the length $L$ with exponents $\eta= 2$ and $\eta = 2 n/\alpha$ for $ n \in \mathbb{Z}^+$ 
\cite{Calabrese2009,calabrese2010universal, cardy2010unusual,igloi2008finite,calabrese2010corrections} 
(see discussion in section \ref{sec-JK}). 
However, the FSE in the other models or in disjointed intervals are unknown (see Fig. \ref{fig-fig1} (b)), which are 
also challenging to be calculated by the JK approach \cite{gu2003entanglement,
gu2005ground,tu2014lattice,gong2017entanglement,devakul2015early}. 
Great endeavor has been made trying to explore this FSE 
in disjointed intervals \cite{alba2011entanglement,fagotti2010entanglement,ruggiero2018entanglement}, and failed to find 
some universal scaling behaviors in them. 

This work aims to explore the scaling law of the above FSE in multiple intervals (see Fig. \ref{fig-fig1} (b)) in free 
Fermion ($c = \bar{c} =1$) and Boson ($c = \bar{c} =1/2$) models, in which the correlators exhibit 
some kind of scaling laws under uniform dilation, such as $\langle \phi(\lambda x) \phi(\lambda y) \rangle = 
\lambda^{-\nu} \langle \phi(x) \phi(y)\rangle$. This feature can give rise to scaling law in 
$\Delta_A^\alpha$ if it is a function of these correlators, which has not yet been unveiled in the previous literature.
Let us denote $A = \cup_i A_i$ to be the jointed structure of some disjointed intervals $A_i$ and $\lambda A$ 
(with $\lambda \in \mathbb{Z}^+$) denotes its uniform  dilation. The set of open ends are denoted as $\partial 
A = \cup_i \partial A_i$. The key result of this work in the large size limit can be formulated as
\begin{equation}
	\Delta_{\lambda A}^\alpha = \Delta^\alpha_A \lambda^{-\eta} \mathcal{B}(A, \lambda),
	\label{eq-DeltaA}
\end{equation}
where $|\mathcal{B}(A, \lambda)| \le 1$ is a bounded function and $\eta = \text{min}(2, 2/\alpha)$ when 
$\alpha < 10$. We find that only the edge modes of the correlation matrix with wave functions localized near 
the open ends contribute to the R\'enyi entropy and its FSE. We confirm Eq. \ref{eq-DeltaA} in both free Fermion and free 
Boson models. Our results may shed new insight into the FSE of R\'enyi entropy in multiple intervals in the other many-body 
systems.

This work is organized as following. In section \ref{sec-XY}, we present the XY model, in which the properties of the 
correlation function is discussed in details. These correlation functions are essential for the scaling laws of the 
entanglement entropy. We will show that the correlation functions in the gapped and gapless phases are totally different. 
In section \ref{sec-JK}, we will discuss the major results by JK. In section \ref{sec-TwoFSE}, two different FSEs
are defined in disjointed intervals, and their features in two and three intervals are discussed. At the end
of this section, the results in the gapped phases, which are trivial, will also be briefly discussed. In section \ref{sec-Con},
we conclude our results. In section \ref{sec-appendix}, we show that the entropy defined in this way is well-defined. 

\begin{figure}[t]
    \centering
    \includegraphics[width=0.48\textwidth]{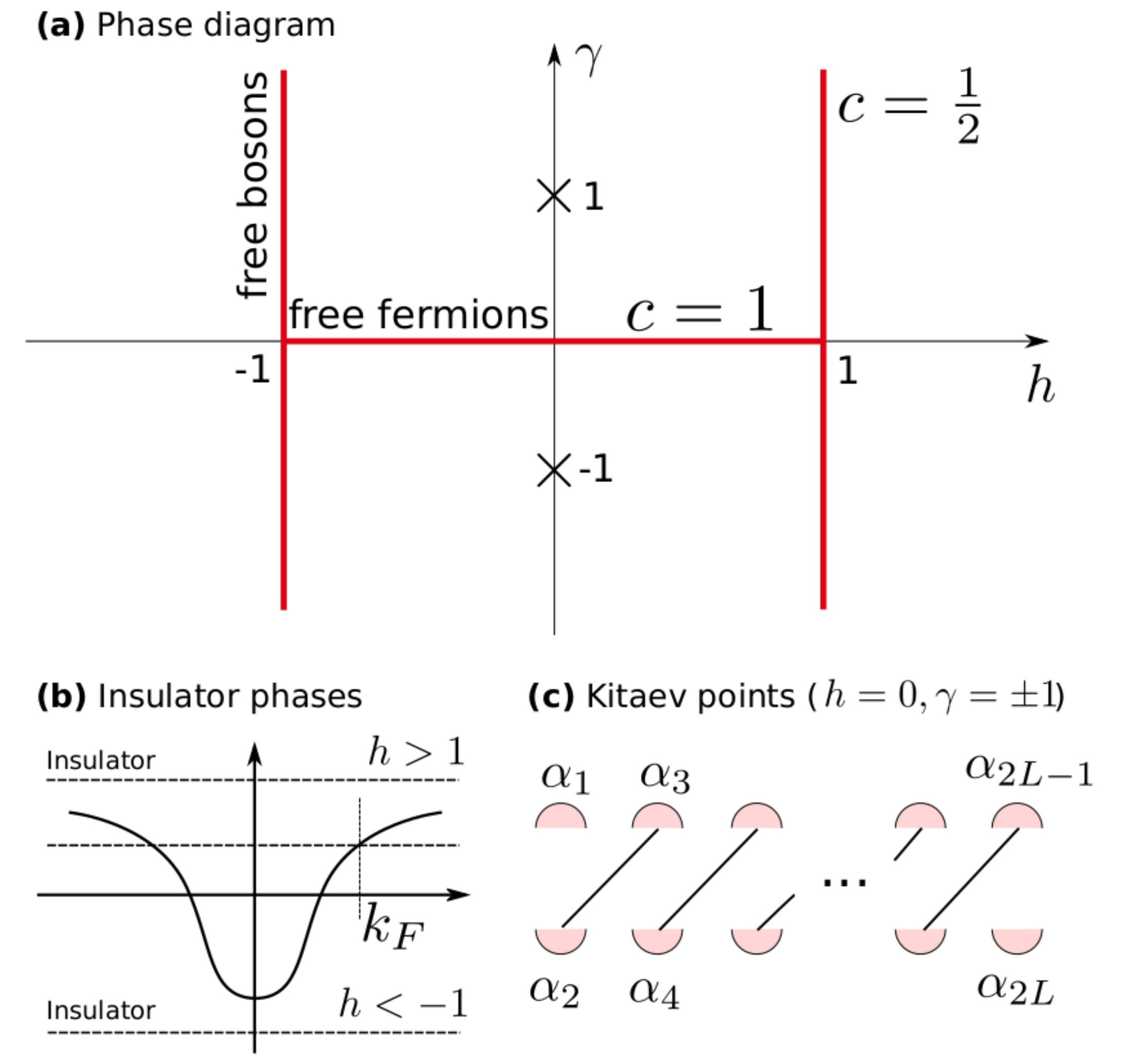}
    \caption{ (a) Phase diagram of the transverse XY model. The thick lines correspond to the gapless phase with $c = \bar{c} = 1$
	for free Fermions and $c =\bar{c} = 1/2$ for free Bosons. 
	(b) Insulator phases with $\gamma=0$ and $|h| > 1$ and gapless phase with Fermi points $\cos(k_F) = h$ when $|h| < 1$. 
	(c) The special points with $\gamma =\pm 1$ and $h = 0$. In the Majorana Fermion representation, this model is decoupled into paired Majorana fermions (represented by the hemicycles), with $\alpha_1$ and $\alpha_{2L}$ unpaired, giving rise to degenerate zero modes. This special case has been studied by Kitaev \cite{Kitaev2001unapired}. The insulator phases for $|h|> 1$ and $\gamma =0$ in (b) and the Kitaev points in (c) have zero range correlation with $\mathcal{G}(x) = 0$. }
    \label{fig-fig2}
\end{figure}

\begin{figure}[t]
    \centering
    \includegraphics[width=0.46\textwidth]{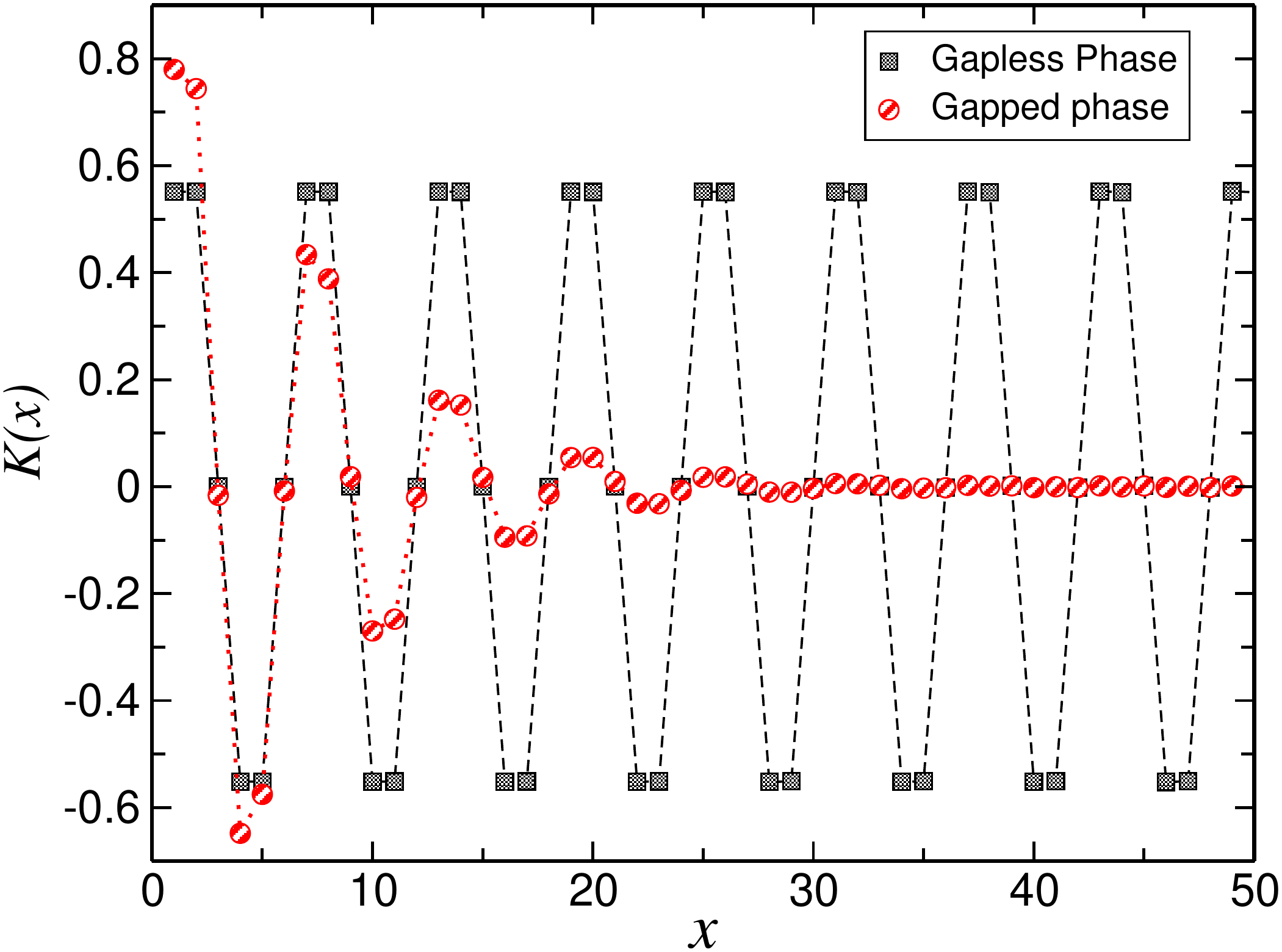}
	\caption{$K(x)$ in the gapped phase and gapless phase. In the gapped phase, we have choen $\gamma = 0.4$ and $h = 0.5$; while 
	in the gapless phase, we used $\gamma = 0.0$ and $h = 0.5$. Only values at $x \in \mathbb{Z}$ are plotted.}
	\label{fig-fig3}
\end{figure}

\begin{figure}[t]
    \centering
    \includegraphics[width=0.48\textwidth]{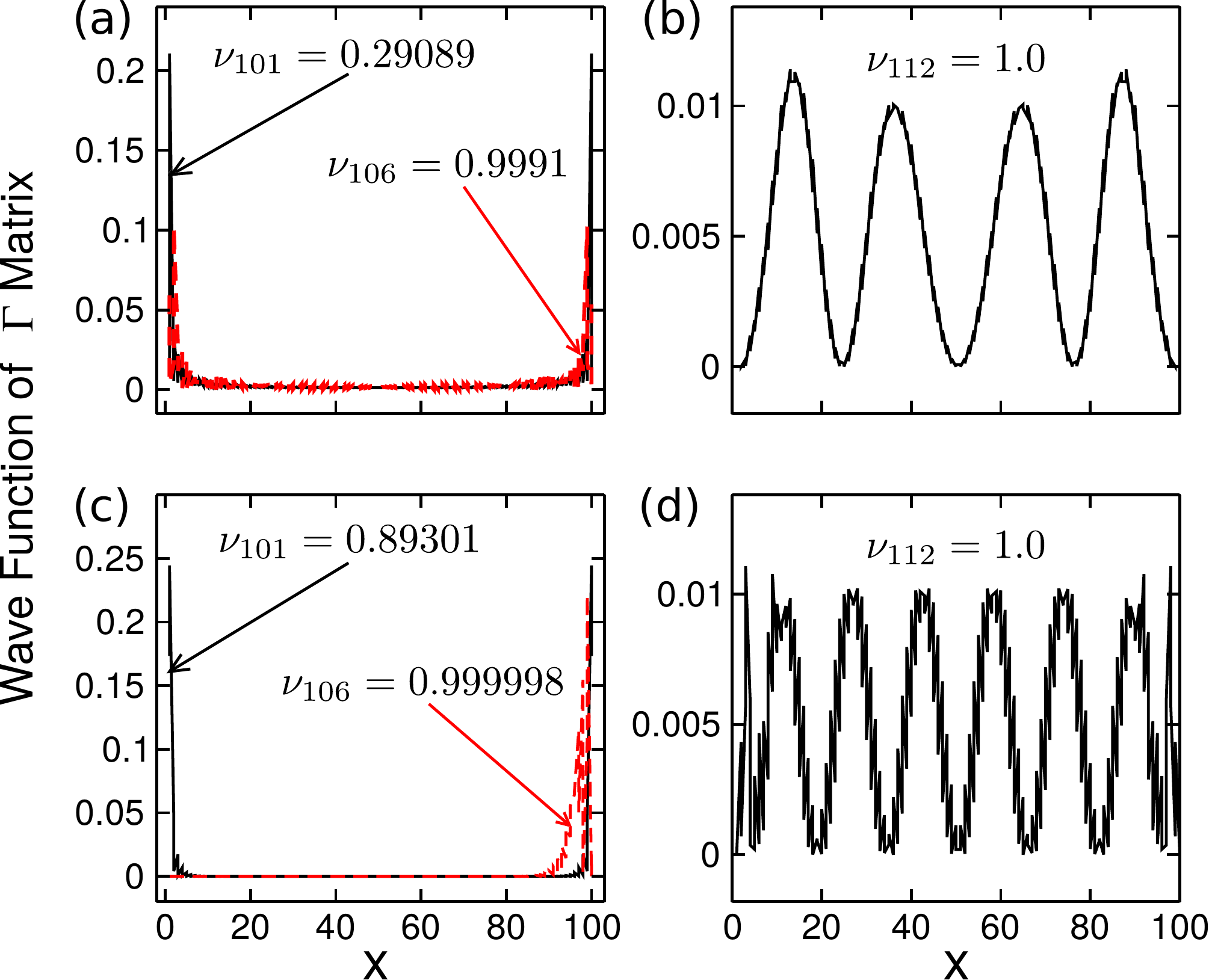}
	\caption{Wave functions of the correlation matrix $i\Gamma$ in a single interval with $L=100$. (a) and (b) show the wave function in the gapless phase ($\gamma =1.36$ and $h=1.0$) for $\nu_{101}$, $\nu_{106}$ and $\nu_{112}$ (using $\nu_{100+l} = -\nu_{100- l}$ from the particle-hole symmetry of $i\Gamma$); and (c) and (d) show the results for $\gamma = 1.36$ and $h=1.5$ in the gapped phase for $\nu_{101}$, $\nu_{106}$ and $\nu_{112}$. Here $i\Gamma$ is a $2L \times 2L$ matrix, thus we have in totally $2L = 200$  eigenvalues.}
	\label{fig-fig4}
\end{figure}

\section{XY spin chain}
\label{sec-XY}

We illustrated the above conclusion using the following exact solvable one dimensional XY spin chain 
\begin{equation}                                                                                                                        
    H = \sum_i (\frac{1+\gamma}{2}) s_i^x s_{i+1}^x + ( \frac{1 - \gamma}{2}) s_i^y s_{i+1}^y + h s_i^z,
\end{equation}
where $s_i^\alpha$ ($\alpha = x, y, z$) are Pauli matrices and $h$ is the transverse Zeeman field. After 
a Jordan-Wigner transformation by assigning Fermion operators $c_i$ and $c_i^\dagger$ to each site, it is
mapped to a free Fermion model as
\begin{equation}
H = -\sum_i c_i^\dagger c_{i+1} + \gamma c_i^\dagger c_{i+1}^\dagger + \text{h.c.} + h(1-2 c_i^\dagger c_i),
	\label{eq-H}
\end{equation}
with excitation gap 
\begin{equation}
\epsilon_k = \sqrt{(\cos(k) - h)^2 + \gamma^2 \sin(k)^2}.
\end{equation}
The phase boundary is determined by $\epsilon_k = 0$, which yields three phase boundaries in Fig. \ref{fig-fig2}. When 
$\gamma \ne 0$, we have $|h| =1$; and when $\gamma = 0$, we have $|h| \le 1$. The phase transition in
this model is characterized by $\mathbb{Z}_2$ symmetry breaking \cite{sachdev2007quantum}. We choose this model 
for the sake that the two gapless boundaries correspond to free Fermions and free Bosons, respectively (see Fig. 
\ref{fig-fig2} (a)), thus this model automatically yields the physics in these two distinct free particles. 

The density matrix of $A = \cup_i A_i$ can be calculated exactly using the same approach as that used in a single interval, for the 
reason that the density matrix of several disjointed intervals can be expressed as \cite{vidal2003entanglement,fagotti2010entanglement}
\begin{eqnarray}
	\rho_A \propto \exp(H_A), \quad H_A = {i \alpha^T W \alpha \over 4}, \quad \tanh{W \over 2} = \Gamma,
	\label{eq-rhoA}
\end{eqnarray}
based on the Majorana operators $\alpha_{2l-1}=(\prod_{m<l} \sigma_m^z )\sigma_{l}^x$ and $\alpha_{2l}=(\prod_{m<l} 
\sigma_m^z )\sigma_{l}^y$. Here, $\Gamma$ is a skew matrix with entries given by 
\begin{eqnarray}
	\Gamma_{i,i+x} && = -i(\langle	\alpha_{i} \alpha_{i+x}\rangle - \delta_{x0}) = \begin{pmatrix}	
		0 & \mathcal{G}(x)  \\                                                                                               
		- \mathcal{G}(-x)   & 0                
	\end{pmatrix}, 
	\label{eq-Gamma}
\end{eqnarray}
where \cite{vidal2003entanglement,latorre2003ground,jin2004quantum}
\begin{eqnarray}
\mathcal{G}(x) = \int_{-\pi}^{\pi}\frac{\gamma \sin(k) \sin(k x)-e_k \cos(k x)}{2\pi \epsilon_k} dk,
	\label{eq-gx}
\end{eqnarray}
with $e_k = h - \cos(k)$. This integral determines all the properties of the R\'enyi entropy. 
It has a number of salient features \cite{latorre2003ground}. In the gapless phases, it decays algebraically 
as
\begin{equation}
	\mathcal{G}(x) = {K(x) \over x},
\end{equation}
where $K(x)$ is a bounded oscillating function. Specifically, we find that: 

(I) For free Fermions with $\gamma = 0$ and $|h| \le 1$, we have
\begin{equation}
	K(x) = {(2 \sin(x\arccos(|h|)) - \sin(\pi x) ) \over \pi}.
\end{equation}
When $h = \pm 1$, the spectra is gapless with quadratic dispersion 
as $E_k \propto k^2$ and $\mathcal{G}(x) = 0$, which violate conformal symmetry. In the fully gapped phase with $|h| > 1$, 
we always have $\mathcal{G}(x) = 0$ for the reason of a vacuum state or a fully filled state (see Fig. \ref{fig-fig2} (b)), 
hence $\Gamma$ is always equal to zero (see Eq. \ref{eq-Gamma}). 

(II) For free Bosons with $|h| = 1$ and $\gamma \ne 0$, we have 
\begin{equation}
K(x) \simeq -2 \text{sign} (\gamma) (1 - \cos(\pi x)),
\end{equation}
which is long-range correlated. The above two $K(x)$ are bounded functions, that is, 
$|K(x)| \le C$ for some positive constant $C$; and $\mathcal{G}(x)$ decays according to $1/x$ besides 
their oscillating behaviors. This long-range correlator is essential for the logarithm relation of the
entanglement entropy, as shown in Eq. \ref{eq-SALog} and Eq. \ref{eq-SA}.  

(III) In the gapped phases, $\mathcal{G}(x)$ is short-range correlated with an exponential 
decaying behavior. Based on Eq. \ref{eq-gx}, we can even show at the Kitaev points with $h=0$ and $\gamma = \pm 1$ 
(see Fig. \ref{fig-fig2} (c)), $\mathcal{G}(x)$ is zero range correlated since 
\begin{equation}
K(x) = \sin(\pi x) = 0, \quad x \in \mathbb{Z},
	\label{eq-kxgapped}
\end{equation} 
which can be understood that in these two points, $\alpha_{2i}$ and $\alpha_{2i+1}$ are paired, leaving 
only $\alpha_1$ and $\alpha_{2L}$ to be the dangling operators left out from the Hamiltonian. 
In the gapped phases with finite energy gap, by expanding $\epsilon_k = a + b k^2$, where $a = |h \pm 1|$ 
and $b = (|h\pm 1| + \gamma^2)/2|h\pm 1|$ for the energy gap at $k=0$ ($-$) or $\pi$ ($+$), we obtain 
\begin{equation}
\mathcal{G}(x) \sim e^{-|x|/\xi},
\end{equation}
where the decay length $\xi = |\gamma| /(\sqrt{2} |h \pm 1|)$. It implies of the area law \cite{Brandao2015},
as shown in Eq. \ref{eq-Area}. In Fig. \ref{fig-fig3}, we plot the results of $K(x)$ in the gapped and gapless phases, 
showing of excellent agreement with the above analysis. In the gapped phase, $K(x)$ will always vanished at large
$x \in \mathbb{Z}$. 

\begin{figure}[t]
    \centering
    \includegraphics[width=0.48\textwidth]{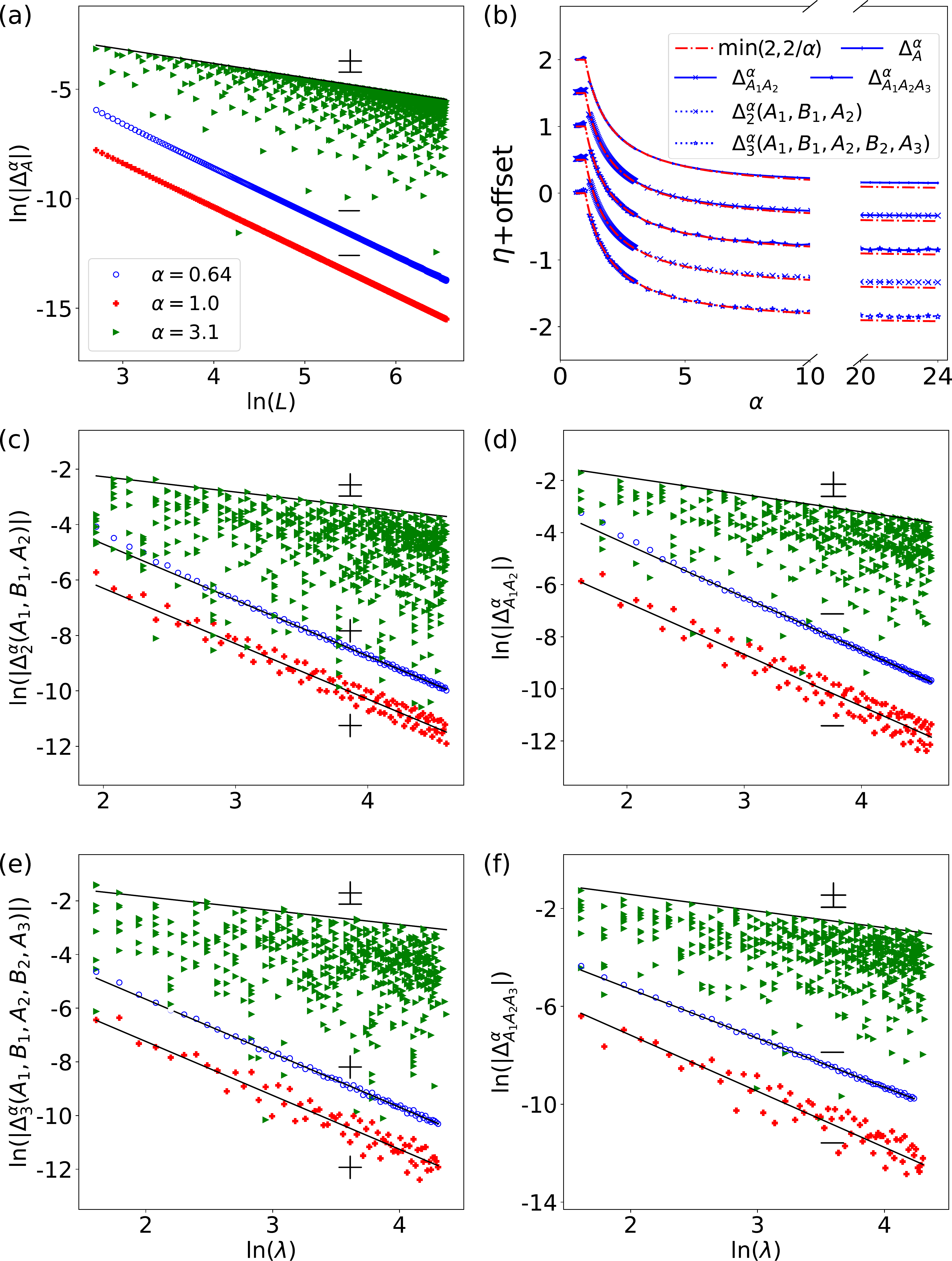}
	\caption{FSE of $\Delta_A^\alpha$ for the free Fermion model at $\gamma = 0$ and $|h| =0.6$. (a) The FSE in
	a single interval; (c) - (f) show the FSE of the entropy for the two different definitions (see text) based on
	two and three disjointed intervals $\lambda A = (A_1,B_1,A_2)=(\lambda,3\lambda,2\lambda)$ (c -d) and 
	$\lambda A = (A_1,B_1,A_2,B_2,A_3)=(\lambda,2\lambda,\lambda,2\lambda,4\lambda)$ (e -f). The plus and minus 
	signs next to each line indicate the sign of this FSE, thus $\pm$ corresponds to the oscillation 
	behavior with period $d = \pi/k_F$ (with $k_F = \arccos(|h|)$) arising from $\mathcal{B}(A, \lambda)$. 
	(b) The fitted values of $\eta$ for these five cases, which for sake of convenient 
	are offset by 0.5; otherwise, they will collapse to the same curve given by $\eta = \text{min}(2, 2/\alpha)$ 
	when $\alpha$ is not large enough. The regime for $\alpha \sim 1$ can not be fitted well using Eq. \ref{eq-DeltaA} 
	from the cancellation effect in Eq. \ref{eq-Cab}. }
	\label{fig-fig5}
\end{figure}

\begin{figure}[t]
	\centering
	\includegraphics[width=0.48\textwidth]{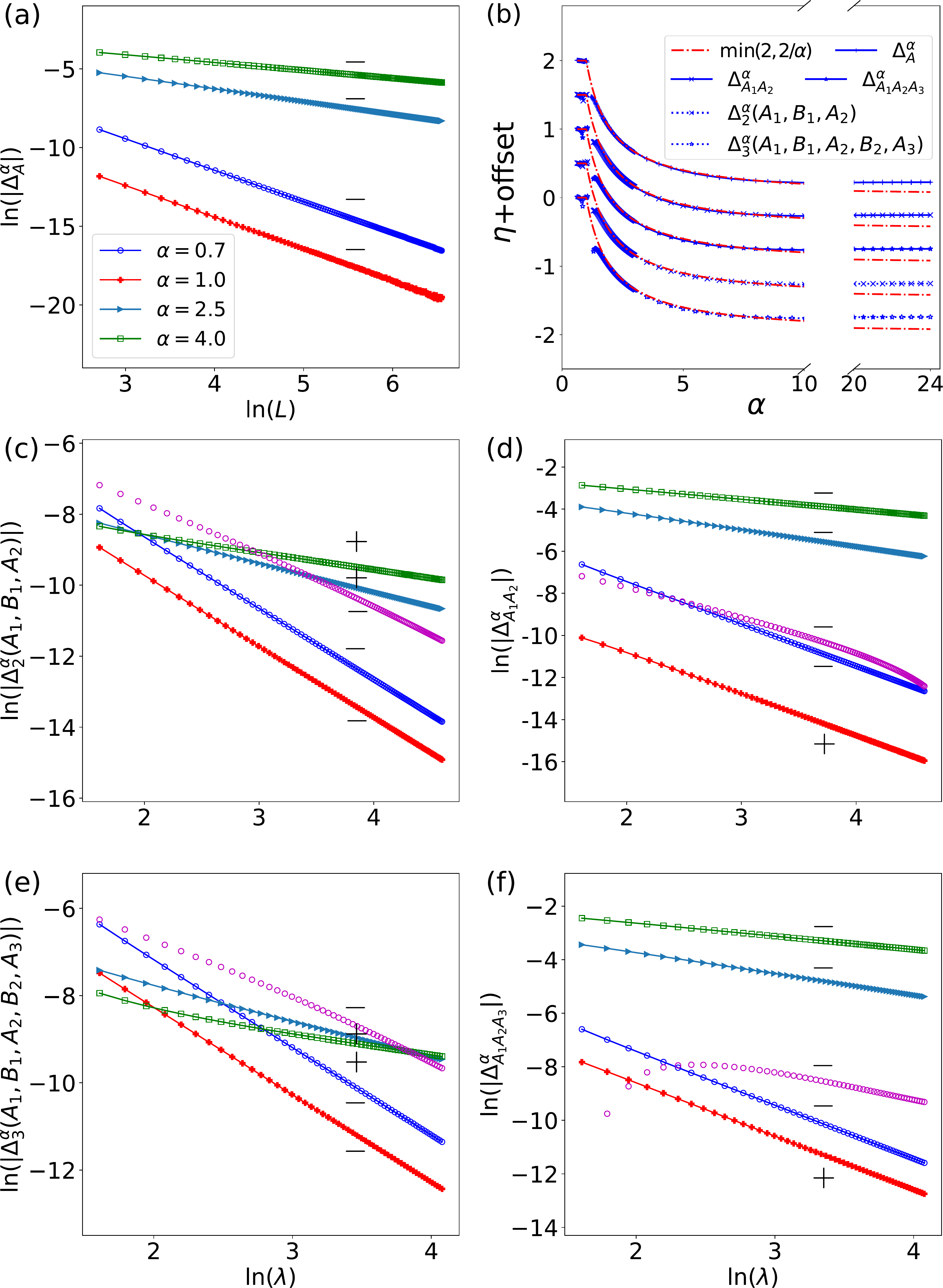}
    \caption{FSE of $\Delta_A^\alpha$ for the free Boson model at $\gamma =1.36$ and $h = 1$. The meaning of each curve
	are the same as that in Fig. \ref{fig-fig5}; however, since $k_F =\pi$ as shown from the correlator $\mathcal{G}(x)$,
	oscillation of the FSE is absent, and the scaling law of Eq. \ref{eq-DeltaA} can be well reproduced. For comparison, we 
	also show the condition of non-uniform dilation using $(A_1,B_1,A_2)=(28, \lambda, 24)$ and $(A_1,B_1,A_2,B_2,A_3)=
	(18,\lambda,12,\lambda, 23)$ with open circles, which violate Eq. \ref{eq-Cab} obviously. }
	\label{fig-fig6}
\end{figure}

Eq. \ref{eq-rhoA} is essential to calculate the density matrix of $\rho_A$ and its eigenvalues numerically. 
Noticed that $W$ and $\Gamma$ are real 
skew matrices, and can be solved by an orthogonal matrix, then we have $\rho_A = \prod_l \otimes \rho_l$, with 
$\rho_l = \text{diag} ({1 - \nu_l \over 2}, {1 + \nu_l \over 2})$, with eigenvalues as $\lambda = 
\prod_{s_l = 1, -1}({1 + s_l \nu_l \over 2})$. By definition we have \cite{ares2014excited}
\begin{equation}
    S_A^\alpha = {1 \over 1 - \alpha} \sum_l \log_2(({1 + \nu_l \over 2})^\alpha + ({1 - \nu_l \over 2})^\alpha),
\end{equation}
which is reduced to the Shannon entropy when $\alpha \rightarrow 1$, with $S_A  = - \sum_{l=1}^{L}
(\frac{1 \pm \nu_l}{2}) \log_2(\frac{1\pm \nu_l}{2})$. Furthermore, when $\alpha$ is large enough, saturation of
the R\'enyi entropy happens, with $S_A^{\infty} = -\sum_{\nu_l > 0} \log_2 ({1 + \nu_l \over 2})$. The eigenvalues and
eigenvectors of the Hermite operator $i\Gamma$ in Fig. \ref{fig-fig4} show that only the modes localized at
the edges have contribution to the R\'enyi entropy, while the extended modes with $\nu_l \rightarrow \pm 1$ will not. 
This is expected, since for the extended modes with vanished amplitudes at the open ends $\partial A$, the coupling 
between regime $A$ with its complement $\bar{A}$ is vanishing small; however for the localized edge modes, the
coupling is strong. This is also the essential origin of the area law quoted above. 

\section{FSE by Jin-Korepin}
\label{sec-JK}

The FSE of this entropy is defined as the difference between the exact (numerical) R\'enyi entropy 
and the prediction from CFT \cite{laflorencie2006boundary,Calabrese2009,calabrese2011entanglement,xavier2012finite}. In a single interval with $\gamma = 0$ for free Fermions, $\Delta_A^\alpha$ 
was obtained by the JK approach \cite{jin2004quantum}. To the leading term \cite{Calabrese2009,
calabrese2010universal}
\begin{equation}
	\Delta_A^\alpha = {\mathcal{A}_1 \over (2L |\sin(k_F)|)^2} + {\mathcal{A}_2 \over (2L |\sin(k_F)|)^{2 \over \alpha}} + \mathcal{O}(L^{-{2n \over \alpha}}),
	\label{eq-Cab}
\end{equation}
where 
\begin{equation}
	\mathcal{A}_1 = {(12(3\alpha^2-7) + (49-\alpha^2)\sin^2(k_F)) (1+\alpha) \over (285\alpha^3)},
\end{equation}
and
\begin{equation}
\mathcal{A}_2 = {2 Q \cos(2k_F L) \over (1-\alpha)}, 
\end{equation}
with $Q = \Gamma(1/2+1/(2\alpha))^2 / \Gamma(1/2 - 1/(2\alpha))^2$, and $k_F = \arccos(|h|)$ is the Fermi momentum.
The next leading terms correspond to $n > 1$. Note that the second term oscillate periodically with spatial period 
$d = \pi/k_F$, reflecting the coupling between the scatterings near the two Fermi momenta $\pm k_F$. This result shows
that the first term is irrelevant when $\alpha > 1$ with $\mathcal{B}=\cos(2k_F L)$, and the second term is 
irrelevant when $\alpha< 1$ with $\mathcal{B} = 1$, while both terms are important near the Shannon entropy with 
$\alpha \sim 1$. Thus $\eta = \text{min}(2, 2/\alpha)$. Our data in Fig. \ref{fig-fig5} (a) - (b) show excellent 
agreement with this prediction. One should be noticed that $\mathcal{A}_1$ and $\mathcal{A}_2$ may have similar 
amplitudes but opposite signs near $\alpha \sim 1$ with a proper choice of $k_F$, which may yield strong cancellation 
between them, thus it can not be fitted well using Eq. \ref{eq-DeltaA} at the regime with $\alpha \sim 1$ in 
Fig. \ref{fig-fig5} (b).

The following section will generalize the results in Eq. \ref{eq-Cab} to disjointed intervals, showing of great
similarity between them. In the disjointed intervals, the FSE is much more complicated, and two FSEs --- the extrinsic
and intrinsic FSEs --- should be defined, all of which have similar scaling laws under uniform dilation, including the 
basic feature of the bounded function $\mathcal{B}(L)$.

\section{Two FSEs in disjointed intervals}
\label{sec-TwoFSE}

To characterize the FSE in multiple intervals, we need to define two more different FSEs, beside of $\Delta_A^\alpha$ 
in a single interval. To this end, we first consider the FSE in two intervals $A_1$, $A_2$ separated by $B_1$ 
(see configuration in Fig. \ref{fig-fig1} (b)), which reads as
\begin{eqnarray}
	S_{A_1A_2}^\alpha = && S_{A_1B_1 A_2}^\alpha - S_{A_1B_1}^\alpha - S_{B_1 A_2}^\alpha + S_{A_1}^\alpha + \nonumber \\
					    && S_{B_1}^\alpha  
	+ S_{A_2}^\alpha  + \Delta_2^\alpha(A_1, B_1, A_2).
	\label{eq-SA2}
\end{eqnarray}
This definition is well-defined (see appendix \ref{sec-appendix}).
The right hand side contains the entropies in all possible single intervals, while the left hand side is the 
entropy of two disjointed intervals \cite{carrasco2017duality,alba2010entanglement,fagotti2010entanglement,facchi2008entanglement}.
Here we have introduced $\Delta_2^\alpha$ to accounts for the difference between the left and right hand sides, making it to
be an exact identity. When the sizes of $A_1$, $B_1$ and $A_2$ approaches infinity, we expect that
\begin{equation}
	\lim_{L_{A_1, A_2, B_1} \rightarrow \infty} \Delta_2^\alpha(A_1, B_1, A_2) = 0.
\end{equation}
Since it is an extrinsic effect to force the above identity, it is termed as the extrinsic FSE 
of the disjointed intervals. Meanwhile, we can write 
\begin{equation}
S_{A_1A_2}^\alpha = S_{A_1A_2}^{\alpha, \text{cft}} + \Delta_{A_1A_2}^\alpha
\end{equation}
using the definition in Eq. \ref{eq-SA}, where $S_{A_1A_2}^{\alpha, \text{cft}}$ is the entropy from CFT that can be 
found in Refs. [\onlinecite{coser2014renyi, calabrese2011entanglement}]. This FSE is an intrinsic effect, not related to
the identity above, it is termed as the intrinsic FSE of the disjointed intervals. 
By definition, we also expect this FSE is vanished when the intervals and its separation approach infinity. 

The expression of $S_{A_1A_2}^{\alpha, \text{cft}}$ may also be inferred from Eq. \ref{eq-SA2}, assuming of negligible 
FSE and $S_A^\alpha$ in a single interval given by Eq. \ref{eq-SA}. Thus we have an identity between all FSEs as
\begin{eqnarray}
	\Delta_{A_1A_2}^\alpha = && \Delta_{A_1B_1 A_2}^\alpha - \Delta_{A_1B_1}^\alpha - \Delta_{B_1 A_2}^\alpha + \nonumber \\
						   && \Delta_{A_1}^\alpha + \Delta_{B_1}^\alpha  + \Delta_{A_2}^\alpha  + \Delta_2^\alpha(A_1, B_1, A_2).
	\label{eq-Delta2}
\end{eqnarray}
From the fact that CFT is analytically exact when the system size is large enough and that the correlation
between the two intervals decreases with the increasing of separation, we expect all $\Delta^\alpha$ 
in Eq. \ref{eq-Delta2} approach zero at large separation according to $\Delta^\alpha \propto 1/L^\eta$.
This limit can be used to extract the non-universal constant of $s_0$ in Eq. \ref{eq-SA}. In the gapless 
phase, all these FSEs are in the same order of magnitude, thus all of them are important; they reflect the 
FSE of $A$ from different aspects. 

This definition can be easily generalized to three (or many) disjointed intervals using the finding from the CFT that
\begin{eqnarray}
S_{A_1A_2A_3}^\alpha && = S_{A_1B_1A_2B_2A_3}^\alpha - S_{A_1B_1A_2B_2}^\alpha  - S_{B_1A_2B_2A_3}^\alpha \nonumber  \\ 
	&& + \cdots +  S_{B_1}^\alpha + S_{A_2}^\alpha + S_{B_2}^\alpha + S_{A_3}^\alpha   + \Delta_3^\alpha,
	\label{eq-SA3}
\end{eqnarray}
where $\Delta_3^\alpha = \Delta_3^\alpha(A_1,B_1, A_2, B_2, A_3)$ is the extrinsic FSE. This expression can also be obtained
from Eq. \ref{eq-SA2}, assuming of $A_2$ to be a union of two disjointed intervals (see section \ref{sec-appendix}). 
This identity has the same structure as Eq. \ref{eq-SA2}. Similarly, we define 
\begin{equation}
S_{A_1A_2A_3}^\alpha =  S_{A_1A_2A_3}^{\alpha, \text{cft}} + \Delta_{A_1A_2A_3}^\alpha,
\end{equation}
where $\Delta_{A_1A_2A_3}^\alpha$ is the intrinsic 
FSE of the three intervals. We can find an identity between all FSEs exactly the same as Eq. \ref{eq-Delta2}. Thus 
we see that the FSE can be fully characterized by these two FSEs in many intervals. In our numerical simulation, we
can extract these two FSEs, and discuss its effect under uniform dilation.

We present the data for these 
$\Delta^\alpha$ in Fig. \ref{fig-fig5} (c) - (f) under uniform dilation for free Fermions with $c = \bar{c} = 1$ (see 
Fig. \ref{fig-fig2} (a)). We find 
a strong oscillation of $\Delta_A^\alpha$ for $\alpha > 1$ in (c) - (f),  which is consistent with Eq. \ref{eq-Cab} with a 
somewhat modified $\mathcal{A}_1$ and $\mathcal{A}_2$; unfortunately, these values can not be determined analytically. 
We find that when $\alpha < 1$, the $\mathcal{A}_1$ term is always relevant with $\mathcal{B} =1$; while when $\alpha > 1$,
the $\mathcal{A}_2$ term is relevant with $|\mathcal{B}(A, \lambda)|$ 
being a complicated yet non-analytical bounded function. These observations yield the major conclusion of Eq. \ref{eq-DeltaA}. 
In Fig. \ref{fig-fig5} (b), we present the fitted exponent $\eta$ as a function of $\alpha$ for all these 
$\Delta_A^\alpha$, all of which falls to the same expression $\eta = \text{min}(2, 2/\alpha)$ when $\alpha < 10$.
When $\alpha \sim 1$, it may not be well fitted using Eq. \ref{eq-DeltaA} for the same reason of cancellation in 
Eq. \ref{eq-Cab}. Moreover, in Figs. \ref{fig-fig5} (b), saturation of entropy  happens when $\alpha > 10$, 
at which the exponent $\eta$ will deviate from $2/\alpha$ from the unspecified high-order terms $L^{-2n/\alpha}$ ($n>1$) 
in Eq. \ref{eq-Cab}; see \cite{jin2004quantum, Calabrese2009,calabrese2010universal}.

These results can also be found for free Bosons with $c =\bar{c} = 1/2$ in Fig. \ref{fig-fig6}.
However, the correlator $\mathcal{G}(x)$ oscillates with period $k_F = \pi$, thus the oscillation of the FSE 
from $\mathcal{A}_2$ is disappeared and $\mathcal{B}(A, \lambda) = 1$ for all disjointed intervals $A$.
As a result, for all the FSEs in the one, two and three disjointed intervals, all the FSEs decay monotonically
with the increasing of dilation ratio $\lambda$, following the claim of Eq. \ref{eq-DeltaA}. We also show that when the 
dilation is non-uniform, this general relation is failed.  In Fig. \ref{fig-fig6} (b), the summarized $\eta$ are also 
the same as that in Fig. \ref{fig-fig5} (b), showing of the similar cancellation effect prescribed by Eq. \ref{eq-Cab} 
even in multiple intervals, though their analytical expressions are impossible.

Finally, we briefly discuss the FSE in the gapped phases. We find that all the $\Delta_A^\alpha$ decay exponentially 
under dilation in the multiple intervals, with $S_A^\alpha$ satisfies the area law \cite{Brandao2015}. 
In these phases, the correlator 
$\mathcal{G}(x)$ also decays exponentially as a function of $x$. From Fig. \ref{fig-fig4} (c) and (d), we show that 
only the edge modes of $i\Gamma$ contribute to $S_A^\alpha$. Due to the short-range correlation from $\mathcal{G}(x)$, 
the FSEs will be quickly disappeared following $\Delta_A^\alpha \sim e^{-x/\xi}$, where $x$ is the minimal separation
between the open ends in $\partial A$. This result is trivial, thus is not presented in this manuscript. 
For this reason, $\Delta_A^\alpha$ in multiple intervals also exhibit different kinds of scaling laws in the gapped 
and gapless phases, which can be used for the diagnostication of phase transitions \cite{chhajlany2007entanglement,
apollaro2016entanglement,orus2008equivalence,jia2008entanglement,zozulya2009entanglement,lundgren2014momentum, 
dumitrescu2017scaling}.

\section{Conclusion}
\label{sec-Con}

\begin{figure}
	\centering
	\includegraphics[width=0.26\textwidth]{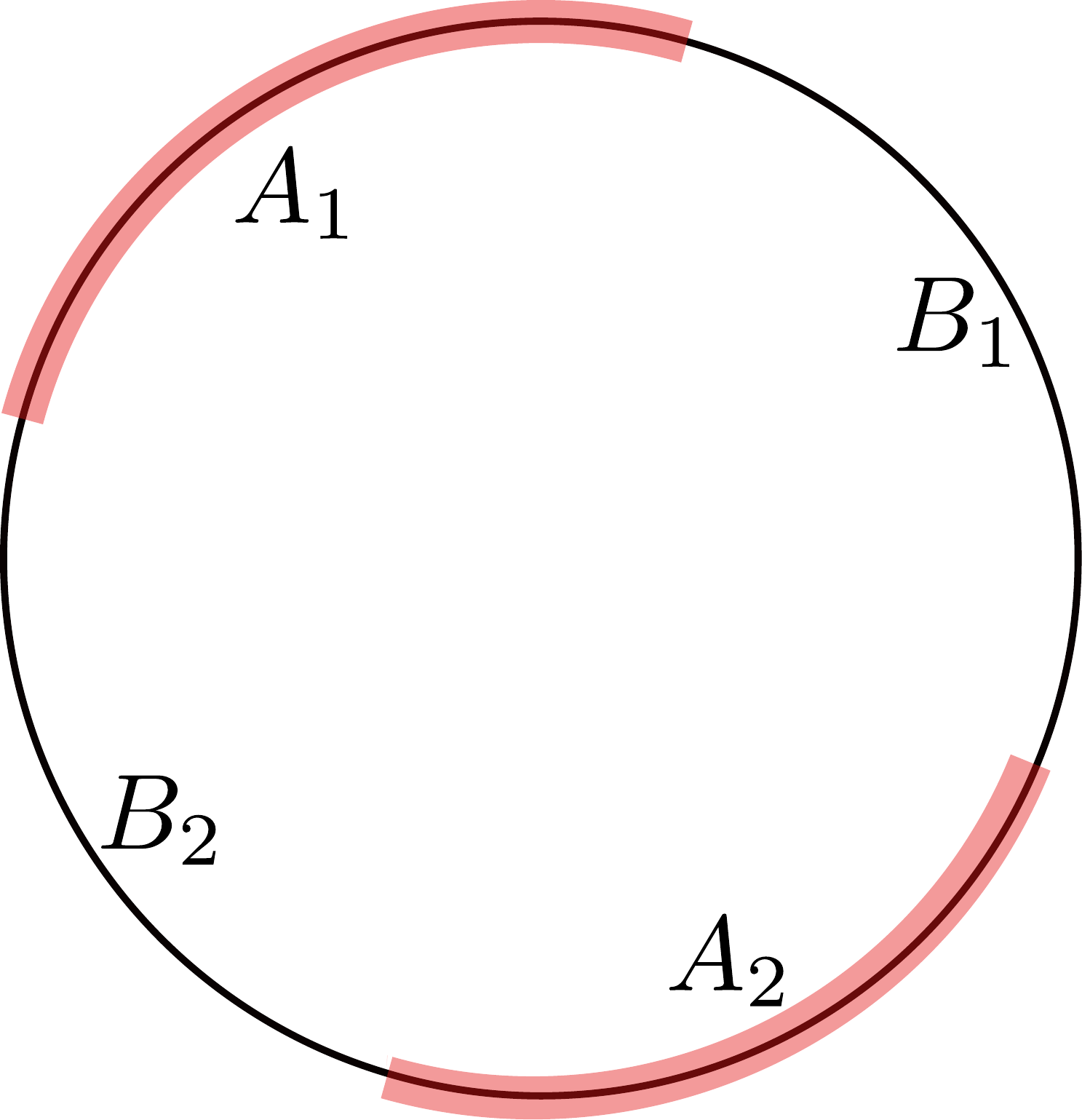}
	\caption{Entanglement entropy in a ring geometry, in which the total entropy of $A = A_1 \cup A_2$ calculated 
	using $A_1$-$B_1$-$A_2$ and $A_1$-$B_2$-$A_2$ will yield the same result.}
	\label{fig-fig7}
\end{figure}

To conclude, we examine the FSE of $\alpha$-R\'enyi entropy in the free Fermion and free Boson models in the 
XY model, which exhibit the same scaling law during uniform dilation that $\Delta_{\lambda A}^\alpha = \lambda^{-\eta}
\Delta_{A}^\alpha \mathcal{B}(A, \lambda)$. 
We find that the regime $\alpha < 1$ and $\alpha > 1$ are described by different relevant terms, 
thus exhibit different scaling behaviors. When $\alpha$ is not large enough, we find $\eta = \text{min} (2, 2/\alpha)$. 
For the Shannon entropy, we thus have $\eta = 2$ exactly. From the correlation matrix $i\Gamma$, we find that only
the edge modes localized at the open ends $\partial A$ contribute to the $\alpha$-R\'enyi entropy as well as its FSE. 
Our results in multiple intervals provide some incisive insight into the entanglement entropy in the many-body system, 
in which the analytical calculation is scarcely possible. Since this FSE is different in the gapped and gapless phases,
the FSE of the disjointed intervals can also be used to characterize this difference and their phase transitions. 

\appendix

\section{The well-defined entropy $S_A$}
\label{sec-appendix}

The formula of entropy in disjointed intervals (see for example in Eq. \ref{eq-SA2} for two intervals) depends only on 
the separation between $A_1$ and $A_2$, which is independent of the other part of the infinity system. The same conclusion 
is for more complicated structures. To understand this, let us assume a ring geometry in Fig. \ref{fig-fig7}, in which $A_1$ and 
$A_2$ are separated by either $B_1$ or $B_1$. We assume that their sizes are large enough, then the FSEs are negligible. We
have two different methods  --- using $A_1$-$B_1$-$A_2$ and $A_1$-$B_2$-$A_2$ --- to accounts for the entropy of $A_1$ and $A_2$, 
that is 
\begin{equation}
	S_{A_1A_2}^\alpha = S_{A_1B_iA_2}^\alpha - S_{A_1B_i}^\alpha - S_{B_iA_2}^\alpha + S_{A_1}^\alpha + S_{A_2}^\alpha + 
	S_{B_i}^\alpha,
\end{equation}
for $i =1$ and $i = 2$. Using the fact that $S_A = S_{\bar{A}}$ by definition, we have $S_{A_1 B_1 A_2}^\alpha = S_{B_2}^\alpha$,
$S_{A_1 B_2 A_2}^\alpha = S_{B_1}^\alpha$,  $S_{A_1 B_2}^\alpha = S_{B_1A_2}^\alpha$, $S_{A_1 B_1}^\alpha = S_{B_2 A_2}^\alpha$, 
and we can show directly that the above two calculations will yield the same result. This method can be generalized to much more
complicated disjointed structures. For this reason, we also expect well-defined FSE of these entropies. 

The above result can also be understood intuitively using the following way. Let us assume that 
$S_{A_1A_2}^\alpha = x_1 S_{A_1B_iA_2}^\alpha  + x_2 S_{A_1B_i}^\alpha  + x_3 S_{B_iA_2}^\alpha + x_4 S_{A_1}^\alpha + x_5 S_{A_2}^\alpha +  x_6 S_{B_i}^\alpha$, where $x_i$ are undetermined coefficients. This definition should satisfy some basic
features. (1) The above two calculations should yield the same result; (2) When the separation $B_1$
and $B_2$ are much larger than the sizes of $A_1$ and $A_2$, we will recover the limit that $S_{A_1A_2}^\alpha = S_{A_1}^\alpha + 
S_{A_2}^\alpha$; (3) This expression has well defined symmetry, that is, $S_{A_1A_2}^\alpha = S_{A_2A_1}^\alpha$. These
three constraints will yield uniquely the above entropy in two disjointed intervals. 

Finally, if we assume that the entropy of two intervals is correct even when $A_2$ is a union of two disjointed intervals. Then 
we assume $A_2 \rightarrow A_2 \cup A_3$, where $A_2$ and $A_3$ are separated by $B_2$. In this way, we will find that 
the right-hand side of Eq. \ref{eq-SA2} is made by disjointed intervals. For instance, $S_{A_1B_1(A_2A_3)}$ is the total
entropy of $A_1\cup B_1 \cup A_2$ and $A_3$, which can be calculated, again, using Eq. \ref{eq-SA2}. Collecting all these
results will yield Eq. \ref{eq-SA3}. In this way, we can derive the expression of entropy in many disjointed intervals. In
the large size limit, the same expression can be found by CFT. This result suggest that 
\begin{equation}
	S_{\lambda A}^\alpha = S_{A}^\alpha + {c + \bar{c} \over 12(1+\alpha)} k \ln \lambda + \Delta_{\lambda A}^\alpha,
\end{equation}
for $k$ disjointed intervals.

{\it Acknowledgments}: This work is supported by the National Key Research and Development Program in China (Grants No. 2017YFA0304504 and No. 2017YFA0304103) and the National Natural Science Foundation of China (NSFC) with No. 11774328. 


%

\end{document}